\documentclass[aps,plb,twocolumn,showpacs,groupedaddress]{revtex4}  
\usepackage{epsfig}                 

\newcommand{\dzero}     {{D\O}}
\newcommand{\uy}        {\ensuremath{ U(1)_{Y}\, }}
\newcommand{\sul}       {\ensuremath{ SU(2)_{L}\,}}
\newcommand{\mzero}     {\ensuremath{ m_0        }}
\newcommand{\mhalf}     {\ensuremath{ m_{1/2}    }}
\newcommand{\tanb}      {\ensuremath{ \tan\beta  }}
\newcommand{\Azero}     {\ensuremath{ A_0        }}
\newcommand{\pt}        {\ensuremath{ p_{T} }}

\newcommand{\Eslash}    {\ensuremath{ E\kern-0.6em\slash }}
\newcommand{\met}       {\ensuremath{ \Eslash_{T}        }}
\newcommand{\metset}    {\ensuremath{ \met/\sqrt{S_{T}} }}
\newcommand{\Rslash}    {\ensuremath{ R\kern-0.6em\slash }}
\newcommand{\rpv}       {\ensuremath{ \Rslash_{p}    }}
\newcommand{\wrpv}      {\ensuremath{ W_{{R\kern-0.46em\slash}_{p}} }}
\newcommand{\lle}       {\ensuremath{ LL\bar{E}}}

\newcommand{\lamoto}    {\ensuremath{ \lambda_{121} }}
\newcommand{\lamott}    {\ensuremath{ \lambda_{122} }}
\newcommand{\lamorr}    {\ensuremath{ \lambda_{133} }}
\newcommand{\mml}       {\ensuremath{ \mu\mu\ell    }}
\newcommand{\eel}       {\ensuremath{ ee\ell        }}
\newcommand{\eet}       {\ensuremath{ ee\tau        }}
\newcommand{\charone}   {\ensuremath{ \tilde{\chi}^{\pm}_1 }}
\newcommand{\chartwo}   {\ensuremath{ \tilde{\chi}^{\pm}_2 }}
\newcommand{\charmpone} {\ensuremath{ \tilde{\chi}^{\mp}_1 }}
\newcommand{\neutone}   {\ensuremath{ \tilde{\chi}^0_1 }}
\newcommand{\neuttwo}   {\ensuremath{ \tilde{\chi}^0_2 }}
\newcommand{\neutthr}   {\ensuremath{ \tilde{\chi}^0_3 }}
\newcommand{\neutfou}   {\ensuremath{ \tilde{\chi}^0_4 }}
\newcommand{\susygen}   {{\sc susygen}}
\newcommand{\suspect}   {{\sc suspect}}
\newcommand{\gauginos}  {{\sc gauginos}}
\newcommand{\pythia}    {{\sc pythia}}
\newcommand{\geant}     {{\sc geant3}}

\begin{document}


  \hspace*{5.2in}\mbox{FERMILAB-PUB-06-089-E}

  \title{Search for {\boldmath$R$}-parity violating supersymmetry via the {\boldmath $LL\bar E$ couplings $\lambda_{121}$, 
    $\lambda_{122}$ or $\lambda_{133}$ in $p \bar p$ collisions at $\sqrt{s}= 1.96$ TeV}}

%
\author{                                                                      
V.M.~Abazov,$^{36}$                                                           
B.~Abbott,$^{76}$                                                             
M.~Abolins,$^{66}$                                                            
B.S.~Acharya,$^{29}$                                                          
M.~Adams,$^{52}$                                                              
T.~Adams,$^{50}$                                                              
M.~Agelou,$^{18}$                                                             
J.-L.~Agram,$^{19}$                                                           
S.H.~Ahn,$^{31}$                                                              
M.~Ahsan,$^{60}$
% ...
\\                                                                            
\vskip 0.30cm                                                                 
\centerline{(D\O\ Collaboration)}                                             
\vskip 0.30cm                                                                 
}                                                                             
\affiliation{                                                                 
\centerline{$^{1}$Universidad de Buenos Aires, Buenos Aires, Argentina}       
\centerline{$^{2}$LAFEX, Centro Brasileiro de Pesquisas F{\'\i}sicas,         
                  Rio de Janeiro, Brazil}                                     
\centerline{$^{3}$Universidade do Estado do Rio de Janeiro,                   
                  Rio de Janeiro, Brazil}                                     
\centerline{$^{4}$Instituto de F\'{\i}sica Te\'orica, Universidade            
                  Estadual Paulista, S\~ao Paulo, Brazil}                     
\centerline{$^{5}$University of Alberta, Edmonton, Alberta, Canada,           
                  Simon Fraser University, Burnaby, British Columbia, Canada,}
\centerline{York University, Toronto, Ontario, Canada, and                    
                  McGill University, Montreal, Quebec, Canada}                
\centerline{$^{6}$Institute of High Energy Physics, Beijing,                  
                  People's Republic of China}                                 
\centerline{$^{7}$University of Science and Technology of China, Hefei,       
                  People's Republic of China}                                 
\centerline{$^{8}$Universidad de los Andes, Bogot\'{a}, Colombia}             
\centerline{$^{9}$Center for Particle Physics, Charles University,            
                  Prague, Czech Republic}                                     
\centerline{$^{10}$Czech Technical University, Prague, Czech Republic}        
\centerline{$^{11}$Center for Particle Physics, Institute of Physics,         
                   Academy of Sciences of the Czech Republic,                 
                   Prague, Czech Republic}                                    
\centerline{$^{12}$Universidad San Francisco de Quito, Quito, Ecuador}        
\centerline{$^{13}$Laboratoire de Physique Corpusculaire, IN2P3-CNRS,         
                   Universit\'e Blaise Pascal, Clermont-Ferrand, France}      
\centerline{$^{14}$Laboratoire de Physique Subatomique et de Cosmologie,      
                   IN2P3-CNRS, Universite de Grenoble 1, Grenoble, France}    
\centerline{$^{15}$CPPM, IN2P3-CNRS, Universit\'e de la M\'editerran\'ee,     
                   Marseille, France}                                         
\centerline{$^{16}$IN2P3-CNRS, Laboratoire de l'Acc\'el\'erateur              
                   Lin\'eaire, Orsay, France}                                 
\centerline{$^{17}$LPNHE, IN2P3-CNRS, Universit\'es Paris VI and VII,         
                   Paris, France}                                             
\centerline{$^{18}$DAPNIA/Service de Physique des Particules, CEA, Saclay,    
                   France}                                                    
\centerline{$^{19}$IReS, IN2P3-CNRS, Universit\'e Louis Pasteur, Strasbourg,  
                    France, and Universit\'e de Haute Alsace,                 
                    Mulhouse, France}                                         
\centerline{$^{20}$Institut de Physique Nucl\'eaire de Lyon, IN2P3-CNRS,      
                   Universit\'e Claude Bernard, Villeurbanne, France}         
\centerline{$^{21}$III. Physikalisches Institut A, RWTH Aachen,               
                   Aachen, Germany}                                           
\centerline{$^{22}$Physikalisches Institut, Universit{\"a}t Bonn,             
                   Bonn, Germany}                                             
\centerline{$^{23}$Physikalisches Institut, Universit{\"a}t Freiburg,         
                   Freiburg, Germany}                                         
\centerline{$^{24}$Institut f{\"u}r Physik, Universit{\"a}t Mainz,            
                   Mainz, Germany}                                            
\centerline{$^{25}$Ludwig-Maximilians-Universit{\"a}t M{\"u}nchen,            
                   M{\"u}nchen, Germany}                                      
\centerline{$^{26}$Fachbereich Physik, University of Wuppertal,               
                   Wuppertal, Germany}                                        
\centerline{$^{27}$Panjab University, Chandigarh, India}                      
\centerline{$^{28}$Delhi University, Delhi, India}                            
\centerline{$^{29}$Tata Institute of Fundamental Research, Mumbai, India}     
\centerline{$^{30}$University College Dublin, Dublin, Ireland}                
\centerline{$^{31}$Korea Detector Laboratory, Korea University,               
                   Seoul, Korea}                                              
\centerline{$^{32}$SungKyunKwan University, Suwon, Korea}                     
\centerline{$^{33}$CINVESTAV, Mexico City, Mexico}                            
\centerline{$^{34}$FOM-Institute NIKHEF and University of                     
                   Amsterdam/NIKHEF, Amsterdam, The Netherlands}              
\centerline{$^{35}$Radboud University Nijmegen/NIKHEF, Nijmegen, The          
                  Netherlands}                                                
\centerline{$^{36}$Joint Institute for Nuclear Research, Dubna, Russia}       
\centerline{$^{37}$Institute for Theoretical and Experimental Physics,        
                   Moscow, Russia}                                            
\centerline{$^{38}$Moscow State University, Moscow, Russia}                   
\centerline{$^{39}$Institute for High Energy Physics, Protvino, Russia}       
\centerline{$^{40}$Petersburg Nuclear Physics Institute,                      
                   St. Petersburg, Russia}                                    
\centerline{$^{41}$Lund University, Lund, Sweden, Royal Institute of          
                   Technology and Stockholm University, Stockholm,            
                   Sweden, and}                                               
\centerline{Uppsala University, Uppsala, Sweden}                              
\centerline{$^{42}$Physik Institut der Universit{\"a}t Z{\"u}rich,            
                   Z{\"u}rich, Switzerland}                                   
\centerline{$^{43}$Lancaster University, Lancaster, United Kingdom}           
\centerline{$^{44}$Imperial College, London, United Kingdom}                  
\centerline{$^{45}$University of Manchester, Manchester, United Kingdom}      
\centerline{$^{46}$University of Arizona, Tucson, Arizona 85721, USA}         
\centerline{$^{47}$Lawrence Berkeley National Laboratory and University of    
                   California, Berkeley, California 94720, USA}               
\centerline{$^{48}$California State University, Fresno, California 93740, USA}
\centerline{$^{49}$University of California, Riverside, California 92521, USA}
\centerline{$^{50}$Florida State University, Tallahassee, Florida 32306, USA} 
\centerline{$^{51}$Fermi National Accelerator Laboratory,                     
            Batavia, Illinois 60510, USA}                                     
\centerline{$^{52}$University of Illinois at Chicago,                         
            Chicago, Illinois 60607, USA}                                     
\centerline{$^{53}$Northern Illinois University, DeKalb, Illinois 60115, USA} 
\centerline{$^{54}$Northwestern University, Evanston, Illinois 60208, USA}    
\centerline{$^{55}$Indiana University, Bloomington, Indiana 47405, USA}       
\centerline{$^{56}$University of Notre Dame, Notre Dame, Indiana 46556, USA}  
\centerline{$^{57}$Purdue University Calumet, Hammond, Indiana 46323, USA}    
\centerline{$^{58}$Iowa State University, Ames, Iowa 50011, USA}              
\centerline{$^{59}$University of Kansas, Lawrence, Kansas 66045, USA}         
\centerline{$^{60}$Kansas State University, Manhattan, Kansas 66506, USA}     
\centerline{$^{61}$Louisiana Tech University, Ruston, Louisiana 71272, USA}   
\centerline{$^{62}$University of Maryland, College Park, Maryland 20742, USA} 
\centerline{$^{63}$Boston University, Boston, Massachusetts 02215, USA}       
\centerline{$^{64}$Northeastern University, Boston, Massachusetts 02115, USA} 
\centerline{$^{65}$University of Michigan, Ann Arbor, Michigan 48109, USA}    
\centerline{$^{66}$Michigan State University,                                 
            East Lansing, Michigan 48824, USA}                                
\centerline{$^{67}$University of Mississippi,                                 
            University, Mississippi 38677, USA}                               
\centerline{$^{68}$University of Nebraska, Lincoln, Nebraska 68588, USA}      
\centerline{$^{69}$Princeton University, Princeton, New Jersey 08544, USA}    
\centerline{$^{70}$State University of New York, Buffalo, New York 14260, USA}
\centerline{$^{71}$Columbia University, New York, New York 10027, USA}        
\centerline{$^{72}$University of Rochester, Rochester, New York 14627, USA}   
\centerline{$^{73}$State University of New York,                              
            Stony Brook, New York 11794, USA}                                 
\centerline{$^{74}$Brookhaven National Laboratory, Upton, New York 11973, USA}
\centerline{$^{75}$Langston University, Langston, Oklahoma 73050, USA}        
\centerline{$^{76}$University of Oklahoma, Norman, Oklahoma 73019, USA}       
\centerline{$^{77}$Oklahoma State University, Stillwater, Oklahoma 74078, USA}
\centerline{$^{78}$Brown University, Providence, Rhode Island 02912, USA}     
\centerline{$^{79}$University of Texas, Arlington, Texas 76019, USA}          
\centerline{$^{80}$Southern Methodist University, Dallas, Texas 75275, USA}   
\centerline{$^{81}$Rice University, Houston, Texas 77005, USA}                
\centerline{$^{82}$University of Virginia, Charlottesville,                   
            Virginia 22901, USA}                                              
\centerline{$^{83}$University of Washington, Seattle, Washington 98195, USA}  
}                                                                             

  \date{April 30, 2006}
  
  \begin{abstract}
    A search for gaugino pair production with a trilepton signature in the framework of $R$-parity 
    violating supersymmetry via the couplings \lamoto, \lamott, or \lamorr\ is presented. 
    The data, corresponding to an integrated luminosity of $\mathcal{L}\approx360$~pb$^{-1}$, 
    were collected from April 2002 to August 2004 with the \dzero~detector at the Fermilab 
    Tevatron Collider, at a center-of-mass energy of $\sqrt{s}=1.96$~TeV. 
    This analysis considers final states with three charged leptons with the flavor combinations 
    \eel, \mml, and \eet\ ($\ell=e$ or $\mu$). No evidence for supersymmetry is found and limits at 
    the 95\%~confidence level are set on the gaugino pair production cross section and lower bounds 
    on the masses of the lightest neutralino and chargino are derived in two supersymmetric models. 
  \end{abstract}

  \maketitle

  Supersymmetry (SUSY)~\cite{bib:susyorig} predicts the existence of a new particle for each 
  standard model (SM) particle, differing by half a unit in spin but otherwise sharing the same 
  quantum numbers.  The new scalar particles, known as squarks and sleptons, carry baryon ($B$) or 
  lepton ($L$) quantum numbers, potentially leading to interactions violating $B$ or $L$ conservation.
  In the supersymmetric Lagrangian, there is a continuous $R$-invariance, which prevents lepton 
  and baryon number violation, but also prevents gluinos and gravitinos from being massive. 

  In a supergravity scenario, the gravitino will acquire mass through the spontaneous breaking of 
  local SUSY. The SUSY-breaking is then communicated to the so-called observable sector so that, in 
  particular, the gluino acquires its mass~\cite{bib:susybreaking}.  This breaks the continuous 
  $R$-invariance, leaving only a discrete version, which is called $R$-parity~\cite{bib:rpvsusy}. 
  Each particle is characterized by an $R$-parity quantum number defined as $R_p = (-1)^{3B+L+2S}$ 
  ($S$ being the spin), such that SM particles have $R_p = 1$ and SUSY particles $R_p=-1$. 
  The gauge symmetry allows  $R$-parity violating (\rpv) terms to be included in the 
  superpotential~\cite{bib:barbier}. These terms are: 
  \begin{eqnarray}
    \wrpv & = &  + \frac{1}{2} \,\lambda _{ijk}\, L_i L_j \bar{E}_k 
    \,           +     \,\lambda^{\prime}_{ijk}\, L_i Q_j \bar{D}_k  \,+\, \mu_i L_i H_u \nonumber \\
    & &          + \frac{1}{2} \,\lambda^{\prime\prime}_{ijk} \bar{U}_i \bar{D}_j \bar{D}_k 
    \label{eq:superpot}
  \end{eqnarray}
  where $L$ and $Q$ are the lepton and quark SU(2) doublet superfields, while $\bar{E}$, $\bar{U}$, and 
  $\bar{D}$ denote the weak isospin singlet fields and the indices $i,j,k$ refer to the fermion families. 
  The coupling strengths in the trilinear terms are given by the Yukawa coupling constants 
  $\lambda$, $\lambda^{\prime}$ and $\lambda^{\prime\prime}$. Terms appearing in the first line of 
  Eq.~(\ref{eq:superpot}) violate lepton number by one unit, and the last term in the second line 
  leads to baryon number violation. The bilinear term $\mu_i L_i H_u$ mixes lepton and Higgs ($H_u$) 
  superfields. 

\begin{figure}[!h]
    \setlength{\unitlength}{1.0cm}
    \begin{picture}(8.0,4.8)(0.0,0.0)
      \put(-0.1,-0.1) {\epsfig{file=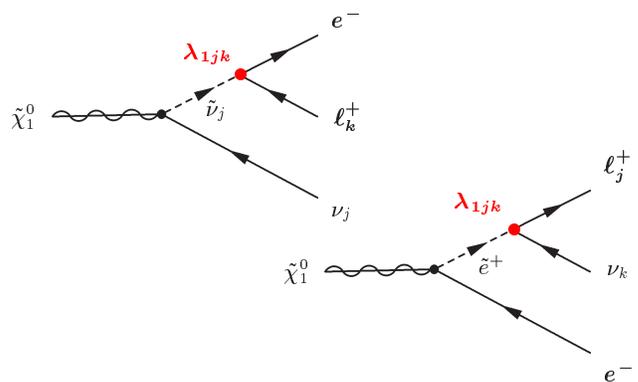, bb= 55 647 295 792, clip= , width=0.96\linewidth}}
    \end{picture}
    \caption{\label{fig:feyndiag}
      Two examples of \rpv-decays of the lightest neutralino via 
      \lle\ couplings $\lambda_{1jk}$. In each decay, two charged 
      leptons and one neutrino are produced.}
  \end{figure}

  \indent This letter reports on a search for chargino and neutralino pair production under the 
  hypothesis that \rpv\ can only occur via a term of the type $\lambda_{ijk}\,L_i L_j \bar{E}_k$.
  A non-zero \rpv\ coupling $\lambda_{ijk}$ thus enables a slepton to decay into a lepton pair, as 
  shown in Fig.~\ref{fig:feyndiag} for the \rpv-decay of the lightest neutralino. The so-called \lle\ 
  couplings $\lambda_{ijk}$ specifically studied here, are \lamoto, \lamott, and \lamorr. 
  One coupling is assumed to be dominant at a time, with any other \rpv-coupling negligibly small.

  The initial state at the Fermilab Tevatron Collider consists of hadrons, so the production of a 
  single SUSY particle could only occur through a trilinear term including at least one baryon field, 
  i.e.\ via $\lambda^{\prime}$ or $\lambda^{\prime\prime}$ terms. Since only the \lle\ term ($\lambda$) 
  is considered here, an assumption is made that SUSY particles are produced pairwise in an $R$-parity 
  conserving process~\cite{bib:allanach}, with \rpv\ manifesting itself in the decay only. Even though 
  direct decays of heavy gauginos ($\tilde{\chi}^0_{2,3,4}$, $\tilde{\chi}^{\pm}_{2}$) are possible, 
  they predominantly cascade decay into the lightest supersymmetric particle (LSP), which in turn 
  decays into SM particles via \rpv. In all scenarios studied here, the lightest neutralino (\neutone) 
  is assumed  to be the LSP. 

  Two SUSY models are investigated. In the minimal supergravity model (mSUGRA)~\cite{bib:sugra}, the 
  universal soft breaking mass parameter for all scalars at the unification scale, \mzero, is set 
  to 100~GeV or 1~TeV. At low \mzero, the stau can be lighter than the second lightest neutralino 
  (\neuttwo) and the lightest chargino (\charone), leading to a larger number of final states with taus. 
  By contrast, a high value of \mzero\ prevents complex cascade decays involving sleptons. 
  The universal trilinear coupling, \Azero, has only a small influence on the gaugino pair production 
  cross section and is set to zero as in the previous Run~I ana\-ly\-sis~\cite{bib:nagy}. Searches for 
  supersymmetric Higgs bosons at LEP~\cite{bib:aleph} imply that $\tanb \leq 2$ is excluded, where 
  \tanb\ is the ratio of the vacuum expectation values of the two neutral Higgs fields. Since the 
  cross section for gaugino pair production increases with increasing \tanb\ due to decreasing masses, 
  a value of $\tanb = 5$ (close to the LEP limit) is chosen to ensure conservative results. A higher 
  value of $\tanb = 20$ is studied exclusively in the \eet\ analysis, because the stau mass decreases 
  with increasing \tanb, leading to an enhanced signal efficiency for this particular analysis. 
  Both signs of the higgsino mixing mass parameter, $\mu$, are considered and the common gaugino 
  mass, \mhalf, is varied. 

  In the specific minimal supersymmetric standard model (MSSM)~\cite{bib:mssm} considered here, 
  heavy squarks and sleptons (1~TeV) are assumed, while the GUT relation between $M_1$ and $M_2$, 
  the masses of the superpartners of the \uy\ and \sul\ gauge bosons, is relaxed. The value of 
  \tanb\ is set to 5, and $M_1$ and $M_2$ are varied independently. The higgsino mixing mass 
  parameter $\mu$ is set to 1~TeV, so that \neutthr, \neutfou, and \chartwo\ are heavy. 

  Within the domain of the SUSY parameters explored in this analysis, pair production of 
  \charone\charmpone\ and \neuttwo\charone\ are the dominant processes, leading to final 
  states with at least four charged leptons and two neutrinos. They come from either the 
  decay of the \neutone, with the lepton flavors depending on $\lambda_{ijk}$, or from 
  cascade decays of \charone\ and \neuttwo.
  The strengths of the couplings are set to 0.01 (\lamoto\ and \lamott) and 0.003 (\lamorr). These values 
  are well below the current limits of \lamoto$\,<\,$0.5,$\,$ \lamott$\,<\;$0.085, and \lamorr$\,<\;$0.005 
  for a slepton mass of 1~TeV, which have been derived from the upper limits \lamoto$\,<\;$0.05,$\,$ 
  \lamott$\,<\;$0.027, and \lamorr$\,<\;$0.0016 obtained for a slepton mass of 100~GeV in 
  Refs.~\cite{bib:barbier,bib:ledsaj}. 
  Additionally, only neutralinos with a decay length of less than 1~cm are considered, which results 
  in a cut-off at low neutralino masses~\cite{bib:dawson}, i.e.\ 30~GeV for \lamoto\ and \lamott, and 
  50~GeV for \lamorr, again for slepton masses of 1~TeV.
  As the \neutone\ can be light, the leptons can have small transverse (w.r.t.\ the beam axis) 
  momentum and thus be difficult to detect. For this reason, only three charged leptons with the 
  flavor combinations \eel, \mml, or \eet\ ($\ell=e$ or $\mu$) are required.

  The analysis is based on a dataset recorded with the \dzero~detector between April 2002 and August 2004, 
  corresponding to an integrated luminosity of $\mathcal{L}=360\pm 23$~pb$^{-1}$. Previous searches with 
  the hypothesis of a \lle\ coupling have been performed by the \dzero~collaboration with Tevatron Run~I 
  data collected at a center-of-mass energy $\sqrt{s}=1.8$~TeV~\cite{bib:nagy}. 

  The \dzero~detector consists of a central tracking system surrounded by a uranium/liquid-argon sampling 
  calorimeter and a system of muon detectors~\cite{bib:run2det}. Charged particles are reconstructed 
  using multiple layers of silicon detectors, as well as eight double layers of scintillating fibers 
  in the 2~T axial magnetic field of a superconducting solenoid.  The \dzero~calorimeter provides 
  hermetic coverage up to pseudorapidities $|\eta|= |-\ln\,[\tan (\theta/2)]| \approx 4$ in a 
  semi-projective tower geometry with longitudinal segmentation. The polar angle $\theta$ is measured 
  from the geometric center of the detector with respect to the proton-beam direction.
  The muon system covers $|\eta|<2$ and consists of a layer of tracking detectors and scintillation 
  trigger counters in front of 1.8~T toroidal magnets, followed by two more similar layers of detectors 
  outside the toroids~\cite{bib:run2muon}.
 
  Events containing electrons or muons are selected for offline analysis by a real-time three-stage 
  trigger system. A set of single and dilepton triggers is used to tag the presence of electrons 
  or muons based on their characteristic energy deposits in the calorimeter, the presence of 
  high-momentum tracks in the tracking system, and hits in the muon detectors. 

  $R$-parity violating supersymmetry events are modeled using \susygen~\cite{bib:susygen}, with 
  CTEQ5L~\cite{bib:cteq5} parton distribution functions (PDFs). The package \susygen\ is interfaced 
  with the program \suspect~\cite{bib:suspect} for the evolution of masses and couplings from the 
  renormalization group equations. 
  Leading order (LO) cross sections of signal processes, obtained with \susygen, 
  are multiplied by a $K$ factor computed with \gauginos~\cite{bib:gauginos}. 
  Standard model processes are generated using the Monte Carlo (MC) generator \pythia~\cite{bib:pythia}. 
  All MC events are processed through a detailed simulation of the detector geometry and response 
  based on \geant~\cite{bib:geant}. Multiple interactions per crossing as well as detector pile-up 
  are included in the simulations. The SM background predictions are normalized using cross section 
  calculations at next-to-leading order (NLO) and next-to-NLO (for Drell-Yan production) with 
  CTEQ6.1M~PDFs$\;$\cite{bib:cteq6}. 
  Background from multijet production is estimated from data similar to the search samples, however, 
  the lepton identification and isolation criteria are inverted (\eel\ and \eet) or loosened (\mml).
  These samples are scaled at an early stage of the analysis where multijet production still dominates. \\[-0.4ex]

  Electrons are identified based on their characteristic energy deposition in the calorimeter.
  The fraction of energy deposited in the electromagnetic part of the calorimeter and the 
  transverse shower profile inside a cone of radius 
  $\Delta\mathcal{R}=\sqrt{(\Delta\eta)^2+(\Delta\varphi)^2}=0.4$ around the cluster direction 
  are considered (where $\varphi$ is the azimuthal angle). In addition, a track must point to the 
  energy deposition in the calorimeter and its momentum and the calorimeter energy must be consistent 
  with each other. Remaining backgrounds from jets are suppressed based on the track multiplicity 
  within $\Delta \mathcal{R}=0.4$ around the track direction. 

  Muons are reconstructed using track segments in the muon system, and each muon is required 
  to have a matched central track measured with the tracking detectors. Furthermore, muons 
  are required to be isolated in both the tracking detectors and the calorimeter, which is 
  essential for rejecting muons associated with heavy-flavor jets. 
  The sum of the track transverse momenta (\pt) inside a cone of $\Delta\mathcal{R}=0.5$ 
  around the muon direction should be less than 2.5~GeV and less than 6\% of the muon \pt. 
  For the calorimeter isolation, a transverse energy ($E_T$) of less than 2.5~GeV in a hollow 
  cone of $0.1<\Delta\mathcal{R}<0.4$ around the muon direction is required and less than 8\% 
  of the muon's transverse energy should be deposited in the calorimeter inside this hollow cone.
  For both isolation criteria, the \pt\ ($E_T$) of the muon track itself is excluded from the sum.

  Electrons and muons are required to be isolated 
  from each other    ($\Delta \mathcal{R}_{e\mu}>0.2$), 
  among themselves   ($\Delta \mathcal{R}_{ee}>0.4$, $\Delta \mathcal{R}_{\mu\mu}>0.2$), and 
  from hadronic jets ($\Delta \mathcal{R}_{\ell j}>0.5$).

  Taus decaying hadronically ($\tau_{had}$) are detected as narrow, isolated jets with a 
  specific ratio of electromagnetic to hadronic energy. Two neural networks (NN) are used to 
  identify one-prong tau decays according to the calorimeter information: either with no 
  subclusters in the electromagnetic section of the calorimeter ($\pi$-like) or with EM 
  subclusters ($\rho$-like)~\cite{bib:ztt}. Muons misidentified as taus are removed by 
  taking the shower shape of the hadronic cluster into account.

  Jets are defined using an iterative seed-based cone algorithm~\cite{bib:midpoint}, clustering 
  calorimeter energy within $\Delta \mathcal{R}= 0.5$. The jet energy calibration is determined from 
  the transverse momentum balance in photon plus jet events. Missing transverse energy (\met) is 
  calculated as the negative vector sum of energy deposits in the calorimeter cells, taking into 
  account energy corrections for reconstructed electrons, muons, and jets. 

  Electron, muon, and tau reconstruction efficiencies and resolutions are determined using 
  measured $Z$ boson decays. They are parametrized as functions of \pt, $\eta$, and $\phi$ 
  and applied to the simulated MC events. 
  The electron and muon trigger efficiencies are measured in data and translate to signal 
  event trigger efficiencies close to 100\% for \eel\ and \eet, and around 94\% for \mml.

  To achieve the best sensitivity for each \rpv-coupling, three different analyses are used 
  depending on the flavors of the leptons in the final state: \eel, \mml, and \eet\ ($\ell=e$ or $\mu$).
  The criteria are summarized in Table~\ref{tab:selection}. Each analysis requires three 
  identified leptons with minimum transverse momenta $\pt^{\ell i}$. In the \eel\ and \eet\ analyses, 
  the same lepton quality criteria are applied to each lepton, independent of its transverse momentum. 
  The \mml\ analysis, however, uses looser quality criteria for the lowest-\pt\ lepton to increase 
  the selection efficiency. Dielectron and dimuon backgrounds from Drell-Yan, $\Upsilon$, and $Z$ 
  boson production are suppressed using cuts on \met\ and on the invariant dilepton mass 
  $M_{\ell\ell}$ (for the \mml\ and \eet\ analyses).
  All three analyses are optimized separately using SM and signal MC simulations. 

  Cuts~I and~II of the \eel\ analysis (Table~\ref{tab:selection}) are used to select a 
  dielectron control sample for data and MC comparison, while cuts III and IV define the trilepton 
  \eel\ analysis. Cut III requires three leptons to be identified, two of which must be electrons. 
  Cut~IV, the photon conversion veto, which requires that a track associated with an electron has 
  hits in the innermost layers of the silicon detector, is extended to  all identified electrons 
  in an event. Contrary to this, cut~II only applies to the two electrons of the control sample.
  In the \mml\ and \eet\ analyses all cuts presented in Table~\ref{tab:selection} 
  serve as selection cuts for the respective trilepton sample and are applied successively to 
  all MC samples and the data. Two-dimensional cuts in the (\met, $M_{\mu\mu}$) and 
  ($\Delta\varphi(\mu\mu)$, \met) planes are defined in the \mml\ analysis to veto events 
  from $\Upsilon$ and $Z$ boson production.
  In the \eet\ analysis, hadronic tau decays are identified by requiring the transverse energy 
  deposited in a calorimeter cone of radius $\Delta\mathcal{R}=0.5$ to be above 10~GeV and an NN 
  output of more than NN$\,>\,$0.9, corresponding to cut~III in Table~\ref{tab:selection}. 
  To select events with real \met, which is expected due to neutrinos in the final state, a cut on 
  \metset\ is applied, where $S_T$ is the total scalar transverse energy. It allows discrimination 
  against events with fake \met, which may arise through statistical fluctuations in jet energy 
  measurements. 

\begin{figure}[!h]
    \setlength{\unitlength}{1cm}
    \begin{picture}(8.0,19.4)(0.0,0.0)
      \put( 0.1,  0.00) {\epsfig{file=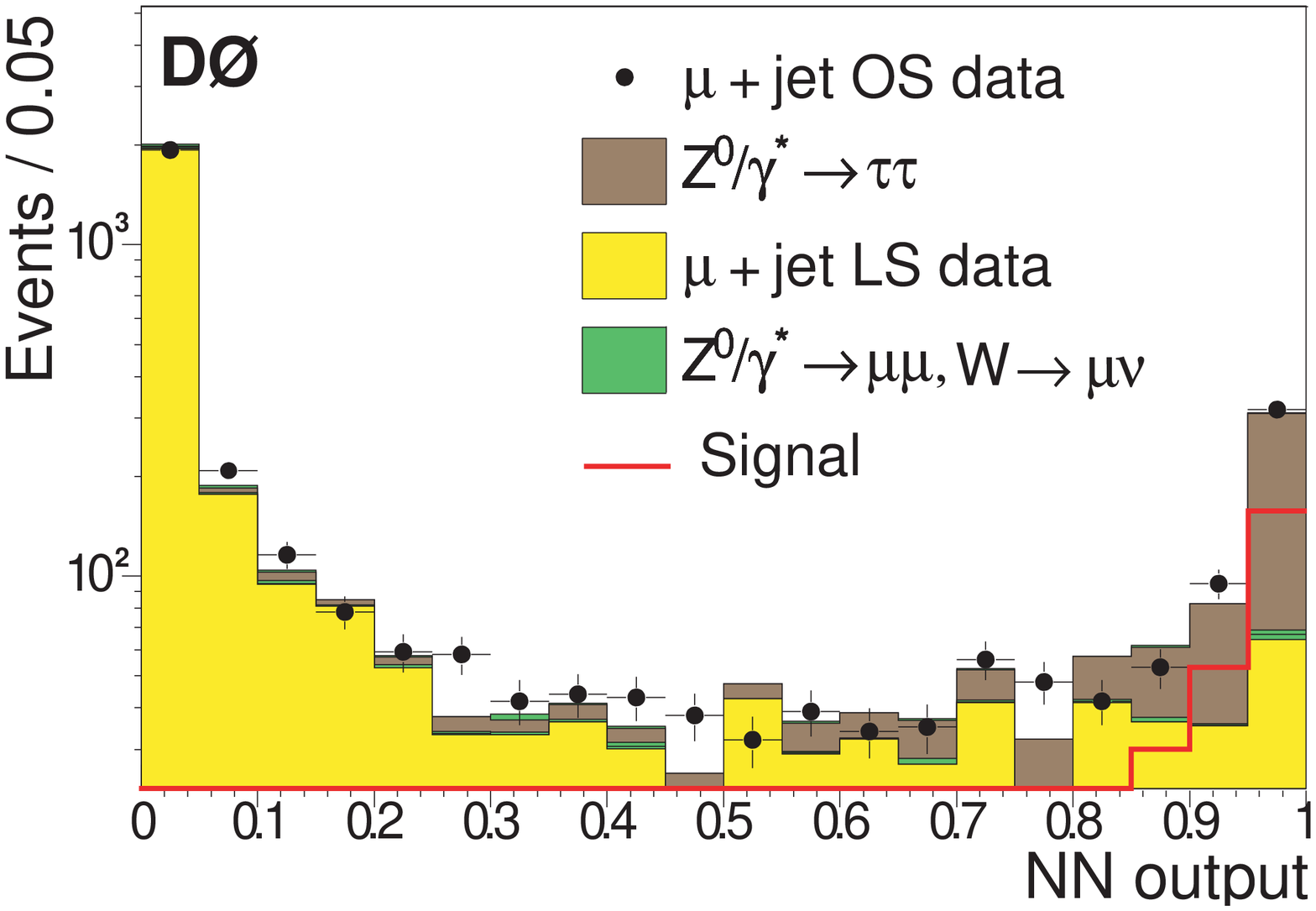, bb= 0 0 606 561, clip= , width=0.93\linewidth}}
      \put( 0.4,  5.50) {\epsfig{file=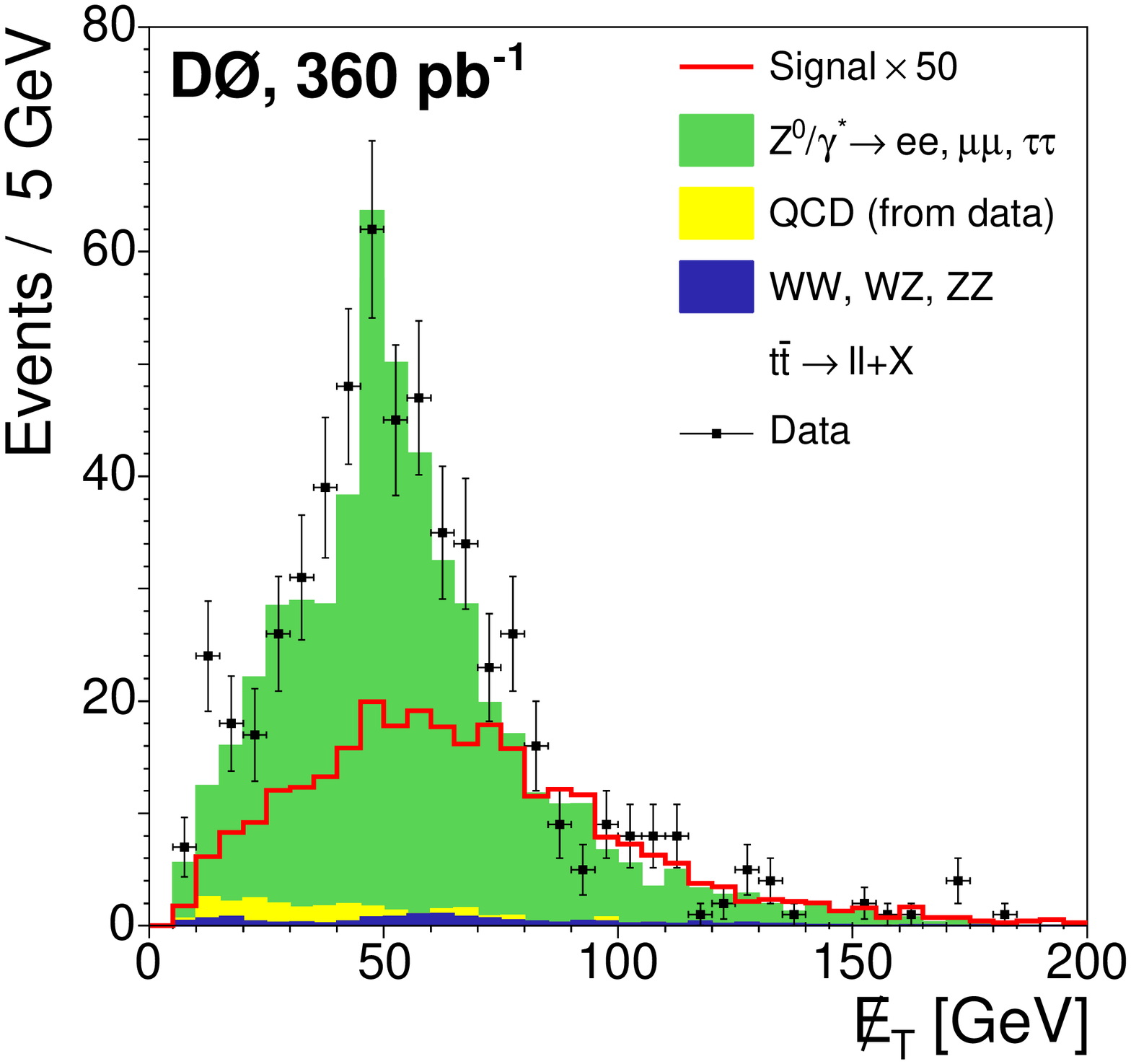, clip= , width=0.85\linewidth}}
      \put( 0.4, 12.45) {\epsfig{file=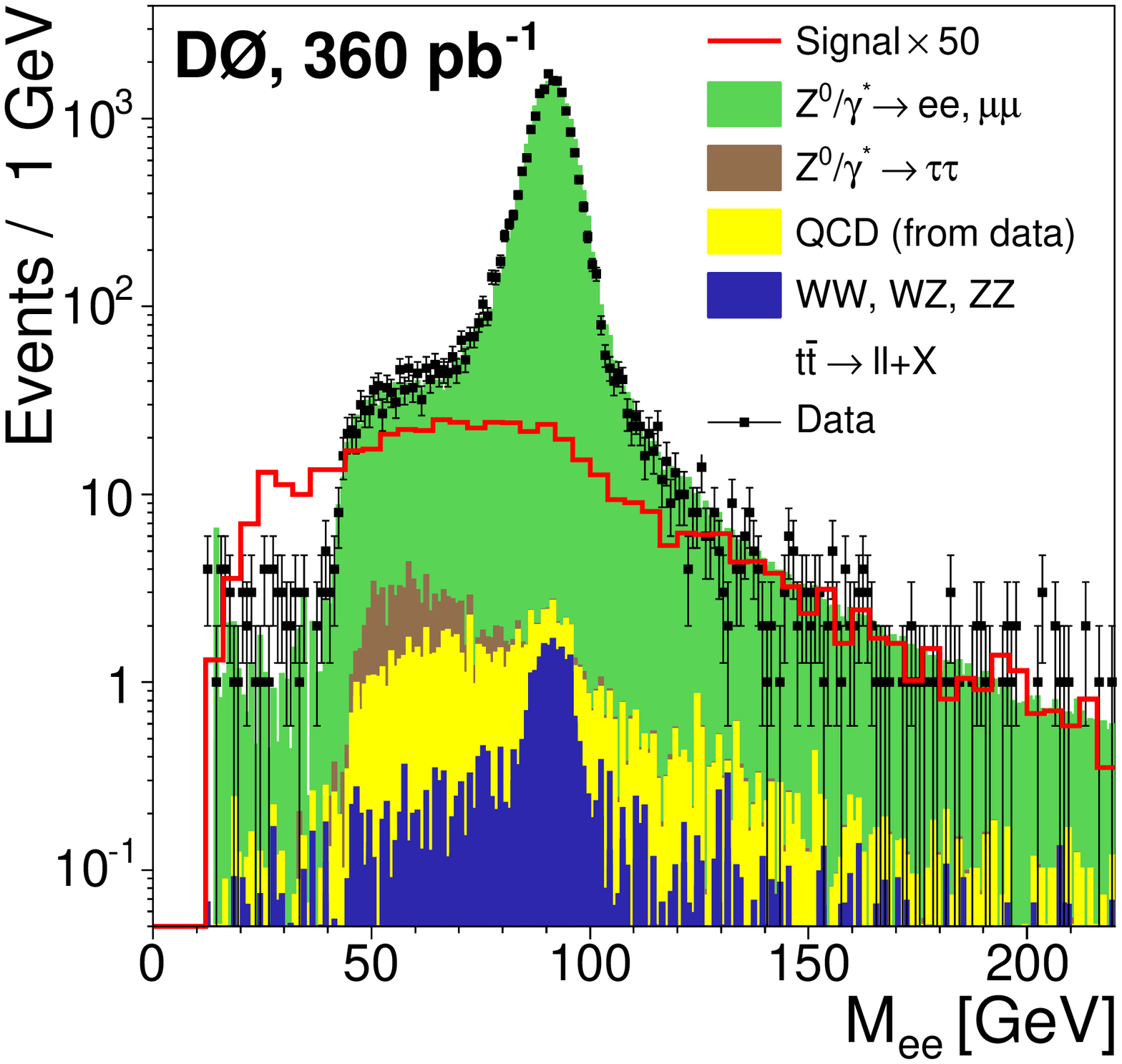, clip= , width=0.85\linewidth}}
      \put(-0.3,  5.0) {(c)}
      \put(-0.3, 11.9) {(b)}
      \put(-0.4, 18.8) {(a)}
    \end{picture}
    \vspace*{-3.5mm}
    \caption{\label{fig:selection}
      (a) The dielectron invariant mass distribution of the \eel\ analysis after 
      cut I, Table \ref{tab:selection}; (b) the \met\ distribution of 
      the \mml\ analysis after cut III, Table \ref{tab:selection}; 
      and (c) the combination of the $\pi$ and $\rho$-like NN outputs of a loose 
      $Z\to\tau\tau \to\tau_{had}\mu$ selection used as the $\tau$ identification 
      criterion in the \eet\ analysis. 
      In (c) like-sign (opposite-sign) $\mu+$jet data is abbreviated LS (OS) 
      and signal refers to the mSUGRA point $\mzero=1$~TeV, $\tanb=5$, $\mu>0$, 
      $\Azero=0$, and $\mhalf=280$~GeV, scaled by a factor of 50 in (a,b) 
      and arbitrarily in (c), details in the text.}
  \end{figure}

  Figure~\ref{fig:selection} shows (a) the dielectron invariant mass in the \eel\ analysis 
  after cut I of Table~\ref{tab:selection}, (b) the missing transverse energy distribution 
  in the \mml\ analysis after cut III of Table~\ref{tab:selection}, and (c) the neural 
  network output for a loose $Z \to \tau\tau \to \tau_{had}\mu$ selection, which is used as an 
  identification criterion for taus in the \eet\ analysis. The $\mu+$jet opposite-sign data sample (OS) 
  represents the control sample, 
  while the $\mu+$jet like-sign data sample (LS) is used to model the multijet background. 
  The different contributions are scaled to the control sample by 
  fitting the $E_T$ spectrum of the tau candidate.
  While in (a,b) the signal is scaled by a factor of 50, an arbitrary scale is used in (c), since the 
  search and control samples are completely independent of each other and no meaningful scale can be 
  defined for the signal contribution w.r.t.\ $Z\to\tau\tau$ or $\mu+$jet data.

  \begin{table}
    \caption{\label{tab:selection} 
      Summary of the selection criteria for \eel, \mml, and \eet\ analyses 
      and numbers of events observed in data and expected from SM background, 
      including statistical and systematic uncertainties.} {\footnotesize
      \begin{ruledtabular}
      	\begin{tabular}{cccc}
      	  \quad & \multicolumn{3}{c} {\bf\boldmath \eel\ ($\ell=e$ or $\mu$) analysis} \\[0.4ex]
      	  \quad &  Cut  &  Data  &  Background \\  
      	  \hline 
      	  I     &  $\pt^{e1}>20$~GeV, $\pt^{e2}>20$~GeV   & $20170$  & $20534\pm 55\pm 1484$ \\   
      	  II    &  $\gamma$-conversion veto (lead. 2 $e$) &          &                       \\
      	  \quad &  and $\met > 15$~GeV                    &  $1247$  &  $1241\pm 21\pm 668$  \\
      	  III   &  $\pt^{\ell1}>20$~GeV, $\pt^{\ell2}>20$~GeV &      &                             \\
      	  \quad &  $\pt^{\ell3}>10$~GeV, at least 2 $e$       &  $5$ & $5.5^{+0.8}_{-0.5}\pm 0.6 $ \\ 
      	  IV    &  $\gamma$-conversion veto (all $e$)         &      &                             \\
      	  \quad &  and $\met > 15$~GeV                        &  $0$ & $0.9^{+0.4}_{-0.1}\pm 0.1 $ \\[0.2ex]
      	  \hline\hline 
      	  \quad &  \multicolumn{3}{c} {\bf\boldmath \mml\ ($\ell=\mu$ or $e$) analysis} \\[0.4ex]
      	  \quad &  Cut  &  Data  &  Background \\
      	  \hline 
      	  I     &  $\pt^{\ell1}>12$~GeV, $\pt^{\ell2}>8$~GeV       & $19283$ & $19588\pm 81\pm 3332$ \\
      	  II    &  $\Delta\varphi(\mu_{i},\met) > 0.1$             & $14918$ & $15275\pm 72\pm 2598$ \\
      	  III   &  $\Upsilon$ and $Z$ veto (\met, $M_{\mu\mu}$) plane &      &                       \\
      	  \quad &  $\Delta\varphi(\mu\mu)<2.53$ for $\met<44$~GeV  &  $564$  &   $506\pm 13\pm   86$ \\
      	  IV    &  $\pt^{\mu3}>4$~GeV or $\pt^e>5$~GeV             &         &                     \\
      	  \quad &  $\Delta\varphi(e,\met) > 0.1$                   &         &                     \\
      	  \quad &  $\sum \pt^{\ell i}> 50$~GeV                     &    $0$  & $0.4\pm 0.1\pm 0.1$ \\[0.2ex]
      	  \hline\hline 
      	  \quad & \multicolumn{3}{c}{\bf\boldmath \eet\ analysis} \\[0.4ex]
      	  \quad &  Cut &   Data &   Background \\
      	  \hline 
      	  I     &  $\pt^{e1}>10$~GeV, $\pt^{e2}>10$~GeV  &          &   \\
      	  \quad &  $M_{ee} > 18$~GeV                     &  $20437$ & $20905\pm 70 \pm 1555$ \\
      	  II    &  $M_{ee} < 80$~GeV                     &   $2831$ &  $2531\pm 32 \pm  329$ \\
      	  III   &  $\tau$: $E_T>10$~GeV, NN$\,>0.9$      &    $16$  &  $ 11.0\pm 2.8\pm 2.0$ \\
      	  IV    &  $\met /\sqrt{S_T}>1.5$ GeV$^{1/2}$    &     $0$  &  $\;1.3\pm 1.7\pm 0.5$ \\[0.4ex]
      	 \end{tabular}
      \end{ruledtabular}
    }
  \end{table}
  
  The number of observed events in data and the expected background from SM processes with its 
  respective statistical and systematic uncertainties are given in Table~\ref{tab:selection}. 
  The multijet background, expressed as a fraction of the SM background, is $11 \pm 7$\%, below 1\% 
  and $15 \pm 15$\% in the \eel, \mml, and \eet\ analyses, respectively.  The number of events 
  observed in data is in good agreement with the expectation from SM processes at all stages of 
  the three analyses. 
  
  The numbers of events expected from SM background and from signal depend on several quantities, 
  each one introducing a systematic uncertainty. The relative uncertainty due to the luminosity 
  measurement is 6.5\%. 
  The relative uncertainty on trigger efficiencies ranges from about 11\% for Drell-Yan (DY) background 
  with low dilepton invariant masses (15~GeV $< M_{\ell\ell}<$ 60~GeV) to about 1\% for the signal. 
  Lepton identification and reconstruction efficiencies give 3\% ($e$), 4\% ($\mu$), and 12\% ($\tau$) 
  per lepton candidate, and the photon conversion veto adds another 0.4\%. The relative systematic 
  uncertainties due to the resolution of the electron or muon energies and \met\ are estimated by 
  varying the resolutions in the MC simulation and are found to be less than 1\% ($e$), 1.5\% ($\mu$), 
  and 2\% (\met). 

  Further systematic uncertainties on the experimental cross section limits concern the 
  theoretical uncertainties on SM background MC cross sections, ranging from 3\% to 17\%, 
  depending on the process, and including PDF uncertainties. Since \pythia\ does not model 
  the $Z$ boson \pt\ accurately, a relative uncertainty of 3\% to 15\%, depending on the 
  dilepton mass, is added for MC Drell-Yan events. 
  The influence of PDF uncertainties on the signal acceptance is estimated to be 4\%.

\begin{figure}[!h]
    \setlength{\unitlength}{1cm}
    \begin{picture}(8.0, 8.0)(0.0,0.0)
      \put(-0.30,-0.1) {\epsfig{file=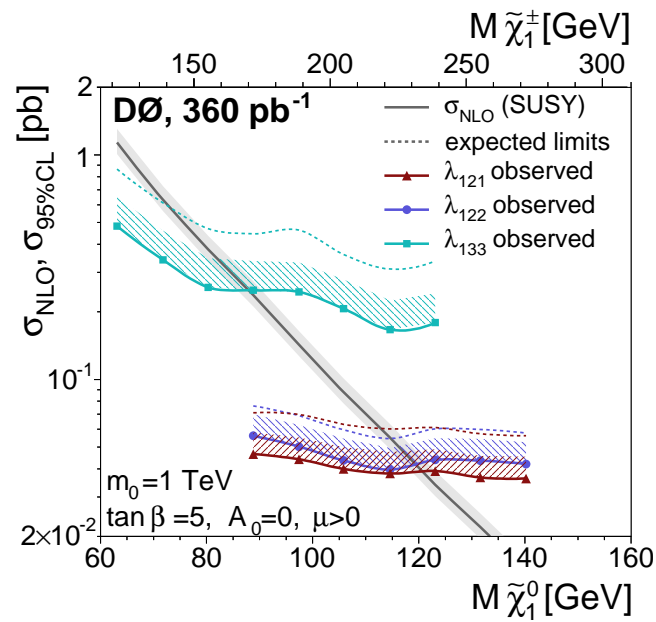, clip= , width=0.99\linewidth}}
    \end{picture}
    \caption{\label{fig:msugra-lim}
      mSUGRA (\mzero$\,=\,$1~TeV, \tanb$\,=\,$5, $\mu>\,$0, \Azero$\,=\,$0): \\
      The $\sigma_{NLO}$ cross section and the $\sigma_{95\% CL}$ limits for 
      the \lamoto, \lamott, and \lamorr\ analyses as functions of the \neutone\ 
      mass (lower horizontal axis) and the \charone\ mass (upper horizontal axis). 
      The exclusion domains, indicated by the hatched regions, lie above the 
      respective observed limit curve. \vspace*{-1mm}}
  \end{figure}

  Theoretical uncertainties on the signal cross sections are due to variations of 
  the renormalization and factorization scales~(5\%), the LO cross section~(2\%), 
  the $K$~factor~(3\%), and the choice of PDF~(9\%). As gaugino pair production mostly 
  proceeds via $s$-channel exchange of virtual $\gamma$, $W$, or $Z$ bosons, the latter 
  uncertainty is deduced from studies of the DY cross section at similar masses. 
  The uncertainty on the DY cross section due to the choice of PDF is estimated to be 6\%, using 
  the CTEQ6.1M\ uncertainty function set~\cite{bib:cteq6}. An additional 3\% is added linearly to 
  account for the lower DY cross section if calculated with CTEQ6~PDFs, compared to its estimation 
  with CTEQ5~PDFs, which are used for the signal MC generation. 
  An additional, conservative, systematic uncertainty of $+10/-0$\% is added to account 
  for the lower LO cross section from \susygen\ compared to the one obtained 
  with \pythia. All of these uncertainties are assumed to be independent, and are added in quadrature.
  The total systematic uncertainty of $-11$\% and $+15$\% is represented by the grey-shaded 
  bands of the signal cross section curve in Fig.~\ref{fig:msugra-lim}. \\

  When setting limits, the \eel, \mml, and \eet\ analyses are combined for each coupling 
  (\lamoto, \lamott, \lamorr) in order to enhance the signal sensitivity. All signal and 
  background samples, as well as the data are processed by all analyses according to the 
  three channels. Events selected in multiple channels are assigned only to the analysis 
  with the largest signal-to-background ratio, and are removed from all other analyses. 
  The percentage of common signal events for any two analyses is less than 13\%, while 
  no common data or SM background events are found. 
  Table~\ref{tab:effis} shows the efficiencies of the analyses for a typical 
  mSUGRA point ($\mzero=1$~TeV, $\tanb=5$, $\mu>0$, $A_0 = 0$, and $\mhalf= 280$ GeV). 
  Correlations between the signal efficiencies in the three channels are taken into 
  account in the calculation of the systematic uncertainties. 
  \begin{table}[!h]
    \vspace*{-3mm}
    \begin{center}
      \caption{\label{tab:effis} 
  	Efficiencies (in \%) of the \eel, \mml, and \eet\ analyses and of 
  	the combined analyses for a typical mSUGRA point ($\mzero=1$~TeV, 
  	$\tanb=5$, $\mu>0$, $\Azero=0$, and $\mhalf=280$~GeV). The first 
	uncertainty is statistical and the second one systematic.}  {\footnotesize
  	\begin{ruledtabular}
  	  \begin{tabular}{lrrr}
  	    Analysis             &  \lamoto\hspace*{6mm}  &       \lamott\hspace*{10mm}        &  \lamorr\hspace*{6mm}\\[0.8ex]
  	    \hline 
  	    $\varepsilon$(\eel)  &  $18.9\pm 0.3\pm 1.2$  &  $ 4.5\;\,\pm 0.2\;\;\pm 0.3\;\;$  &  $2.6\pm 0.2\pm 0.1$ \\
  	    $\varepsilon$(\mml)  &  $ 2.1\pm 0.1\pm 0.3$  &  $16.1\;\,\pm 0.1\;\;\pm 1.9\;\;$  &  $0.8\pm 0.1\pm 0.1$ \\
  	    $\varepsilon$(\eet)  &  $ 1.1\pm 0.1\pm 0.1$  &  $0.23\pm 0.04\pm 0.03$            &  $2.0\pm 0.2\pm 0.2$ \\[0.2ex]
  	    \hline 
  	    \quad                &        \quad           &          \quad                     &        \quad         \\[-2.6ex]
  	    $\varepsilon_{comb}$ &  $22.1\pm 0.3\pm 1.6$  &  $20.8\;\,\pm 0.2\;\,\pm 2.2\;\;$  &  $5.4\pm 0.3\pm 0.4$ \\
  	  \end{tabular}
  	\end{ruledtabular}
  	\vspace*{-3mm}
      }
    \end{center}
  \end{table}

  Since no evidence for gaugino pair production is observed, upper limits on the cross sections are 
  extracted in two models: in mSUGRA (with $\mzero=100$~GeV or 1 TeV, $\tanb=5$ or 20, $\mu>0$, and 
  $A_0=0$) and in an MSSM model assuming no GUT relation between $M_1$ and $M_2$ and assuming heavy 
  squarks and sleptons, i.e. the higgsino mixing mass parameter, $\mu$, and all sfermion masses are 
  set to 1 TeV. Limits are calculated at the 95\%~C.L. using the LEP CL$_S$ method \cite{bib:tlimit} 
  taking into account correlated uncertainties between SM and signal processes. 
  
  For mSUGRA ($\mzero=1$~TeV and $\tanb=5$), the expected and observed cross section limits 
  ($\sigma_{95\% CL}$) are shown in Fig.~\ref{fig:msugra-lim} as functions of the \neutone\ 
  and \charone\ masses. 

  Studies for $\mzero=100$~GeV and $\tanb=5$ and 20 are done for \lamorr. 
  Particularly interesting is the region of high \tanb\ values, where the stau is the 
  next-to-lightest supersymmetric particle. In such a case, decays of SUSY particles into 
  final states with stau leptons can be dominant and consequently increase the efficiency 
  of the \eet\ channel. Lower bounds on the masses of the \neutone\ and the \charone\ are 
  given in Table~\ref{tab:msugra-lim}. 

\begin{figure}[!h]
    \setlength{\unitlength}{1cm}
    \begin{picture}(8.0,7.4)(0.0,0.0)
      \put(-0.30,-0.2) {\epsfig{file=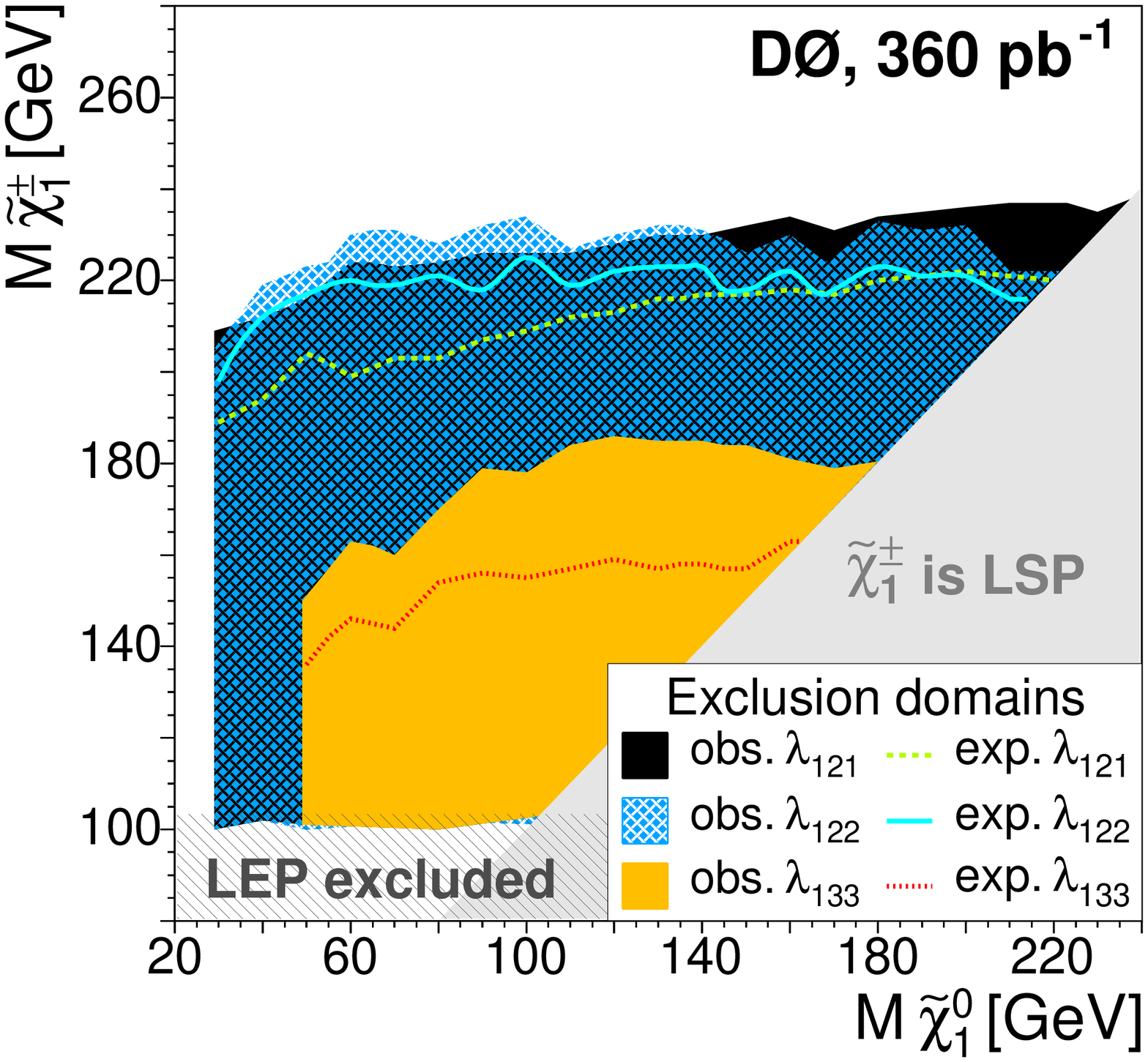, clip= , width=1.00\linewidth}}
    \end{picture}
    \caption{ \label{fig:mssm-lim}
      Observed and expected exclusion domains at the 95\%~C.L. in the 
      (\neutone, \charone) mass plane of the considered MSSM model for 
      the \lamoto, \lamott, and \lamorr\ couplings with their strengths 
      set to 0.01 (\lamoto, \lamott) and 0.003 (\lamorr). \vspace*{-3mm}}
  \end{figure}

  In the MSSM, the exclusion domain is presented in the (\neutone, \charone) mass plane in 
  Fig.~\ref{fig:mssm-lim}. The cut-off of the exclusion domain towards low neutralino masses, 
  i.e. at $m_{\neutone}=30$~GeV for \lamoto\ and \lamott, and at $m_{\neutone}=50$~GeV for 
  \lamorr, is due to the combined effect of the mean decay length of the lightest neutralino 
  (chosen to lie below one cm) and the values of the \lamoto, \lamott, and \lamorr\ couplings. \\

  \begin{table}[!h]
    \caption{\label{tab:msugra-lim} 
      The combined lower limits at the 95\%~C.L. on the masses 
      of \neutone\ and \charone\ (in GeV) obtained using the 
      mSUGRA model with different parameters.} {\footnotesize 
    \begin{ruledtabular}
      \begin{tabular}{lccc}
	Coupling                                & sign($\mu$) & $m(\neutone)$ & $m(\charone)$ \\[1.0ex]
	\hline 
	\lamoto\ ($\mzero=1$~TeV, $\tanb=5$)    &     $>0$    &     119       &    231   \\ 
        \lamott\                                &     $>0$    &     118       &    229   \\
	\lamorr\                                &     $>0$    &      86       &    166   \\[0.4ex]
	\hline 
	\lamoto\ ($\mzero=1$~TeV, $\tanb=5$)    &     $<0$    &     117       &    234   \\         
	\lamott\                                &     $<0$    &     115       &    230   \\[0.4ex]
	\hline 
	\lamorr\ ($\mzero=100$~GeV, $\tanb=5$)  &     $>0$    &     105       &    195   \\
	\lamorr\ ($\mzero=100$~GeV, $\tanb=20$) &     $>0$    &     115       &    217   \\
      \end{tabular}
    \end{ruledtabular}
    }
  \end{table}
  In summary, no evidence for \rpv-SUSY is observed in trilepton events. Upper limits on the 
  chargino and neutralino pair production cross section are set in the case of one dominant  
  coupling: \lamoto, \lamott, or \lamorr. Lower bounds on the masses of the lightest neutralino 
  and the lightest chargino are derived in mSUGRA and in an MSSM scenario with heavy sfermions, 
  but assuming no GUT relation between $M_1$ and $M_2$. All limits significantly improve previous 
  results obtained at LEP~\cite{bib:barbier} and with the \dzero~Run~I dataset~\cite{bib:nagy} 
  and are the most restrictive to date. \\
  
%
We wish to thank M. Klasen for providing us with the \gauginos\ 
package for the calculation of the $K$ factors for the SUSY signal.
We thank the staffs at Fermilab and collaborating institutions, 
and acknowledge support from the 
DOE and NSF (USA);
CEA and CNRS/IN2P3 (France);
FASI, Rosatom and RFBR (Russia);
CAPES, CNPq, FAPERJ, FAPESP and FUNDUNESP (Brazil);
DAE and DST (India);
Colciencias (Colombia);
CONACyT (Mexico);
KRF and KOSEF (Korea);
CONICET and UBACyT (Argentina);
FOM (The Netherlands);
PPARC (United Kingdom);
MSMT (Czech Republic);
CRC Program, CFI, NSERC and WestGrid Project (Canada);
BMBF and DFG (Germany);
SFI (Ireland);
The Swedish Research Council (Sweden);
Research Corporation;
Alexander von Humboldt Foundation;
and the Marie Curie Program.
%

  

\end{document}